# Efficient parameter sensitivity computation for spatially-extended reaction networks

C. Lester,[1, a)] C.A. Yates,[2] and R.E. Baker[1]

[1)]*Mathematical Institute, University of Oxford, Woodstock Road, Oxford, OX2 6GG.*

[2)]*Department of Mathematical Sciences, University of Bath, Claverton Down, Bath, BA2 7AY.*

(Dated: 5 September 2016)

Reaction-diffusion models are widely used to study spatially-extended chemical reaction systems. In order to understand how the dynamics of a reaction-diffusion model are affected by changes in its input parameters, efficient methods for computing parametric sensitivities are required. In this work, we focus on stochastic models of spatially-extended chemical reaction systems that involve partitioning the computational domain into voxels. Parametric sensitivities are often calculated using Monte Carlo techniques that are typically computationally expensive; however, variance reduction techniques can decrease the number of Monte Carlo simulations required. By exploiting the characteristic dynamics of spatially-extended reaction networks, we are able to adapt existing finite difference schemes to robustly estimate parametric sensitivities in a spatially-extended network. We show that algorithmic performance depends on the dynamics of the given network and the choice of summary statistics. We then describe a hybrid technique that dynamically chooses the most appropriate simulation method for the network of interest. Our method is tested for functionality and accuracy in a range of different scenarios.

---

[a)]Electronic mail: lesterc@maths.ox.ac.uk

## I. INTRODUCTION

A wide variety of modeling techniques have been developed to describe spatially-dependent biological phenomena. The dynamics of any given model will necessarily depend on experimentally-derived input parameters. In order to better understand the relationships between input and output variables, it is desirable to perform a parameter sensitivity analysis to understand how changes in model parameters affect the system dynamics.

Recent advances in computational power have made it possible to develop stochastic, individual-based models that enable the study of the behaviour of the individual particles that comprise a biological system of interest[1–3]. These stochastic models explicitly take randomness into account, and have, under a wide variety of circumstances, been shown to be better predictors of model behaviour than corresponding deterministic models[4,5]. In this work, we will focus on voxel-based or lattice-based stochastic models[6]. A volume of interest, $\Omega$, is discretized into a finite number of voxels. The constituents of the model are called particles. Each particle is located within a voxel, and is able to move (diffuse) by transferring into a neighbouring voxel. We further assume that within each voxel the particles are "well-mixed" and can react with one another. This framework is described by the reaction-diffusion master equation (RDME). The RDME has a tractable, analytic solution in only a small number of special cases[6]. In general, due to the high dimensionality of the state space, the RDME is typically analytically intractable and parameter sensitivity analysis must be accomplished through the use of Monte Carlo simulation.

A variety of analytical tools have been developed for performing parameter sensitivity analysis on spatially-homogeneous systems. In this case, the particles are contained within a large, single voxel. The dynamics of such well-mixed systems are typically described by the chemical master equation (CME). Finite difference methods for parameter sensitivity analysis have been developed by Rathinam et al.[7] and Anderson[8]. Exact, likelihood ratio or pathwise methods have been described by Plyasunov and Arkin[9], and Sheppard et al.[10]. Meanwhile, Liao et al.[11] have described a tensor-based method for calculating sensitivities for CME systems.

Methods of efficient exploration of parametric sensitivities in spatially-extended, stochastic systems are less well developed. Mathematically, CME (well-mixed) and RDME (spatially-extended) models are both continuous-time, discrete-state Markov chains. This means that, in principle, any parameter sensitivity analysis method developed for well-mixed systems can be used for spatially-extended systems (see section II A). However, this does not necessarily guarantee that the parameter sensitivity analysis method will be efficient. In this regard, this work makes three contributions. Firstly, we demonstrate that a finite difference scheme can be used to efficiently estimate parametric sensitivities for a spatially-extended model, under a range of circumstances. Secondly, we exploit particular features of the model of interest to describe a novel *grouped-sampling* method. Finally, we implement what we call the *multichoice technique* that dynamically combines different finite difference simulation schemes to efficiently estimate parametric sensitivities.

In section II, we describe the stochastic kinetics described by the RDME, and how these kinetics might be simulated. In section III we describe finite difference methods for parameter sensitivity analysis for spatially homogeneous models. We compare and contrast simulation methods with cases studies in section IV. Grouped-sampling is implemented in section V, and the multichoice method is described in section VI. We conclude in section VII.

## II. STOCHASTIC CHEMICAL KINETICS

We first describe the RDME model[6]. A volume of interest, $\Omega$, is considered. Each particle within $\Omega$ belongs to a particular chemical species: we have $N$ species in total and label the chemical species as $S_1, \ldots, S_N$. In this work, $\Omega$ is assumed to be a volume of dimensions $L \times a \times a$, where $a \ll L$. As such, initially we assume that spatial variability in the distribution of the particles occurs in only the first dimension. We therefore discretize $\Omega$ into $K$ equally sized voxels of dimensions $h \times a \times a$, where $h = L/K$. The voxels are labeled as $V^1, V^2, \ldots, V^K$. Then, $S_i^k$ is used to refer to a particle of chemical species $S_i$ that is located in voxel $V^k$, whilst $X_i^k$ represents the population of $S_i^k$. The population matrix $\boldsymbol{X}$ represents all the chemical populations, and is defined as

$$\boldsymbol{X} = \begin{bmatrix} X_1^1 & \ldots & X_1^K \\ \vdots & & \vdots \\ X_N^1 & \ldots & X_N^K \end{bmatrix}. \tag{1}$$

The particles are able to diffuse (move) within $\Omega$, and can react to change the chemical populations of the system. Particles diffuse by jumping from one voxel into an adjacent voxel, and boundary conditions can be implemented to handle the behaviour at the ends of the domain. Each particle diffuses to a particular neighboring voxel with an average rate of

$$d \equiv \frac{D}{h^2},$$

where $D$ is the diffusion constant[6]. Therefore, with zero-flux boundary conditions, the diffusion of each species $S_i$ can be represented by the collection of events

$$S_i^1 \rightleftharpoons S_i^2 \rightleftharpoons \ldots \rightleftharpoons S_i^{K-1} \rightleftharpoons S_i^K,$$

where $S_i^k \rightleftharpoons S_i^{k+1}$ denotes two events: firstly, diffusion of an $S_i$ particle from voxel $V^k$ to voxel $V^{k+1}$, and, secondly, diffusion of an $S_i$ particle from voxel $V^{k+1}$ to voxel $V^k$.

Reactions describe the changes to the chemical populations of the model, and take place between reactants in the same voxel. Each reaction has two quantities associated with it: firstly, a propensity, that describes the average rate at which the reaction takes place; and, secondly, a stoichiometric vector, that describes how the reaction changes the population levels of the particles. We consider $M$ reaction types, which are labeled as $R_1, \ldots, R_M$. We will assume that each reaction can take place in each voxel, and so we refer to a reaction of type $R_j$ taking place in voxel $V^k$ as $R_j^k$. For further information we refer readers to Erban et al.[6].

## A. Simulation methods

A wide variety of stochastic simulation algorithms (SSAs) are suitable for generating sample paths of RDME models[6]. Perhaps the most widely-used method for producing sample paths according to the RDME is the Gillespie Direct Method (DM)[12]. We use the DM to describe changes to the population matrix (1): the events that change the population vector are given by the set $(\zeta_j)_{j \in \{1,\ldots,J\}}$, so that there are $J$ events in total. The set $(\zeta_j)_{j \in \{1,\ldots,J\}}$ includes the diffusion of particles, and the various reactions that can take place (of which there are $M \cdot K$ possibilities). The propensity of event $\zeta_j$ is given by $\eta_j$, and the stoichiometric matrix as $\kappa_j$. Should event $\zeta_j$ take place at time $t$, the population matrix $\boldsymbol{X}$ is updated as follows:

$$\boldsymbol{X}(t) = \boldsymbol{X}(t-) + \kappa_j. \tag{2}$$

This approach is sometimes known as the "all events method"[13].

The DM is described in algorithm 1. Other simulation methods include the Modified Next Reaction Method (MNRM, described in section II B)[14]. Specialist methods include the Next Subvolume Method[15], and software packages such as URDME[16] have been developed.

## B. The random time change representation

The RDME can also be formulated using the random time change representation, as described by Ethier and Kurtz[17]. The number of times an event $\zeta_j$, with propensity $\eta_j$, fires during the time interval $[0, T]$ is given by the Poisson counting process

$$\mathcal{P}_j \left( \int_0^T \eta_j(\boldsymbol{X}(t)) \mathrm{d}t \right),$$

where $\mathcal{P}_j$ is a unit-rate Poisson process, and $\mathcal{P}_j(t)$ counts the number of arrivals of the unit-rate process within the time interval $(0, t]$. Every time event $\zeta_j$ takes place, the population

---

Algorithm 1: The Gillespie DM. This simulates a single sample path according to the RDME. At each step of the loop, the propensity values are calculated, the time to the next event determined using an exponential variate, and finally the event type chosen. The population values and time are then updated. The loop is repeated until the terminal time is reached.

---

**Require:** initial conditions, $\boldsymbol{X}(0)$, and terminal time, $T$.
 1: set $\boldsymbol{X} \leftarrow \boldsymbol{X}(0)$ and set $t \leftarrow 0$
 2: **loop**
 3:    calculate propensity values $\eta_j$ and set $\eta_0 \leftarrow \sum_{j=1}^{J} \eta_j$
 4:    set $\Delta \leftarrow \mathrm{Exp}(\eta_0)$
 5:    **if** $t + \Delta > T$ **then**
 6:      break
 7:    **end if**
 8:    choose an event $\zeta_k$ to occur: $\zeta_j$ occurs with probability $\eta_j/\eta_0$
 9:    set $\boldsymbol{X} \leftarrow \boldsymbol{X} + \kappa_k$, and set $t \leftarrow t + \Delta$
10: **end loop**

matrix is updated according to equation (2). Therefore, by considering all possible events that might take place over a time interval $(0, T]$, we have the following update formula:

$$\boldsymbol{X}(T) = \boldsymbol{X}(0) + \sum_{j=1}^{J} \mathcal{P}_j \left( \int_0^T \eta_j(\boldsymbol{X}(t)) \mathrm{d}t \right) \cdot \kappa_j. \quad (3)$$

We now show how equation (3) can used to simulate a sample path with the MNRM[14]. If, at time $t$, the system is in state $\boldsymbol{X}(t)$, we need to work out the time until the next event $\Delta$, as well as the particular event which occurs, $\zeta_j$. We determine $\Delta$ by repeatedly performing the following test: suppose event $\zeta_j$ fires next, then what would the putative value of $\Delta$ be? We exhaustively loop through all the events $(\zeta_j)_{j \in \{1,\ldots,J\}}$, to determine putative values for $\Delta$. The event $\zeta_j$ that gives rise to the smallest putative value for $\Delta$ is the one that will fire next. To calculate the value of $\Delta$, for each event $\zeta_j$ we define the following two quantities:

- $P_j = \int_0^t \eta_i(\boldsymbol{X}(t')) \mathrm{d}t'$, the internal or natural time of the event;

- $T_j$, the time of the next arrival in the unit-rate Poisson process $\lambda_j$.

We exhaustively search for the value of $\Delta$ by taking

$$\Delta = \min_j \left( \frac{T_j - P_j}{\eta_j(\boldsymbol{X}(t))} \right), \quad (4)$$

and setting $k$ to be the index where this minimum is obtained. The event $\zeta_k$ now takes place; this simulation method is formally described in algorithm 2.

The DM and MNRM can both be optimized so that the CPU time to produce each individual sample path is kept to a minimum. The focus of this work will, however, be complementary. Instead of minimizing the simulation time required per sample path, we aim to minimize the number of sample paths required to efficiently and accurately estimate parameter sensitivities using the DM and MNRM.

## III. FINITE DIFFERENCE METHODS

In this section we describe how to carry out an efficient parameter sensitivity analysis on a RDME model. In particular, we will assess the effect of a change in the value of an

---

Algorithm 2: The MNRM simulates a single sample path according to the RDME. At each step of the loop, the next event is chosen. The population values and time are then updated. A new random number is generated to replace the one that has just been used to simulate an event, and the loop is repeated until the terminal time is reached.

---

**Require:** initial conditions, $\boldsymbol{X}(0)$ and terminal time, $T$.
 1: set $\boldsymbol{X} \leftarrow \boldsymbol{X}(0)$
 2: set $t \leftarrow 0$
 3: for each $\zeta_j$, set $P_j \leftarrow 0$, and generate $T_j \leftarrow \text{Exp}(1)$
 4: **loop**
 5:    calculate propensity values $\eta_j$ for each $\zeta_j$
 6:    calculate $\Delta_j$ as
$$\Delta_j = \frac{T_j - P_j}{\eta_j}$$
 7:    set $\Delta \leftarrow \min_j \Delta_j$, and $k \leftarrow \operatorname{argmin}_j \Delta_j$
 8:    **if** $t + \Delta > T$ **then**
 9:      break
10:    **end if**
11:    set $\boldsymbol{X} \leftarrow \boldsymbol{X} + \kappa_k$
12:    set $t \leftarrow t + \Delta$
13:    for each $\zeta_j$, set $P_j \leftarrow P_j + \eta_j \cdot \Delta$
14:    set $T_k \leftarrow T_k + \text{Exp}(1)$
15: **end loop**

---

individual input parameter on suitable model summary statistics. If a summary statistic of interest is written as

$$\mathbb{E}\left[f\left(\boldsymbol{X}\right)\right], \tag{5}$$

where $\boldsymbol{X}$ represents a sample path of our RDME model and $f(\boldsymbol{X})$ is a suitable function of interest[1], then we might wish to estimate a partial derivative with respect to a change in the input parameter $A = \alpha$ as

$$\left.\frac{\partial \mathbb{E}\left[f\left(\boldsymbol{X}\right)\right]}{\partial A}\right|_{A=\alpha}. \tag{6}$$

This partial derivative can be estimated for different values of $\alpha$, and choices of parameter $A$. Unsurprisingly, analytic expressions for (6) will not, in general, be obtainable[2], and so stochastic methods must be used to estimate the partial derivative given by (6).

We use a finite difference method to estimate the quantity described by (6). Suppose that

---

[1] For example, $f(\boldsymbol{X})$ could represent a population value at terminal time $T$.
[2] Where the reaction network comprises zero-th and first order reactions only, an analytic solution is available.

Systems $X$ and $Y$ are identical, except that the value of parameter $A$ is perturbed. Pick $\varepsilon \leq 1$, and take $A := \alpha - \varepsilon/2$ in System $X$. For System $Y$, take $A := \alpha + \varepsilon/2$. Then[8],

$$\frac{\partial \mathbb{E}\left[f\left(\boldsymbol{X}\right)\right]}{\partial A} = \frac{\mathbb{E}\left[f\left(\boldsymbol{Y}\right)\right] - \mathbb{E}\left[f\left(\boldsymbol{X}\right)\right]}{\varepsilon} + \mathcal{O}(\varepsilon^2). \tag{7}$$

The centered finite difference approximation[3] is therefore given by $\mathbb{E}\left[f\left(\boldsymbol{X}\right)\right]/\varepsilon$. A first attempt may involve generating sample paths to estimate $\mathbb{E}\left[f\left(\boldsymbol{Y}\right)\right]$ and $\mathbb{E}\left[f\left(\boldsymbol{X}\right)\right]$ independently: but this is usually very inefficient. As such, we explain how to efficiently estimate $\mathbb{E}\left[f\left(\boldsymbol{Y}\right) - f\left(\boldsymbol{X}\right)\right]$ using a variance reduction technique.

We simulate $\mathcal{N}$ pairs of sample paths, which we enumerate as

$$\left\{[\boldsymbol{X},\boldsymbol{Y}]^{(r)}, r = 1,\ldots,\mathcal{N}\right\},$$

and then take $\widehat{\mathbb{Q}}$ as our estimate for $\mathbb{E}\left[f\left(\boldsymbol{Y}\right) - f\left(\boldsymbol{X}\right)\right]$ where $\widehat{\mathbb{Q}}$ is defined as

$$\widehat{\mathbb{Q}} = \frac{1}{\mathcal{N}} \sum_{r=1}^{\mathcal{N}} \left[f\left(\boldsymbol{Y}^{(r)}\right) - f\left(\boldsymbol{X}^{(r)}\right)\right]. \tag{8}$$

The Central Limit Theorem ensures that, in distribution, as $\mathcal{N}$ increases,

$$\frac{1}{\mathcal{N}} \sum_{r=1}^{\mathcal{N}} \left[f\left(\boldsymbol{Y}^{(r)}\right) - f\left(\boldsymbol{X}^{(r)}\right)\right] \to \mathbb{N}\left(\mu, \frac{\sigma^2}{\mathcal{N}}\right),$$

where $\mu$ and $\sigma^2$ are the mean and variance of $[f(\boldsymbol{Y}) - f(\boldsymbol{X})]$. We can therefore construct a confidence interval around our estimator $\widehat{\mathbb{Q}}$ to determine how statistical errors affect our estimate. An approximate 95% confidence interval is typically provided by

$$\left(\widehat{\mathbb{Q}} - 1.96\sqrt{\frac{\sigma^2}{\mathcal{N}}}, \widehat{\mathbb{Q}} + 1.96\sqrt{\frac{\sigma^2}{\mathcal{N}}}\right), \tag{9}$$

where $\sigma^2$ is estimated using the $\mathcal{N}$ sample values of $[f(\boldsymbol{Y}) - f(\boldsymbol{X})]$. To ensure a high degree of statistical accuracy we would like the confidence interval provided in (9) to be small. There are two ways to achieve this: we can either ensure that $\mathcal{N}$ is sufficiently large (i.e. produce many sample paths, leading to relatively long simulation times), or we can use a variance reduction technique to ensure that $\sigma^2$ is relatively small (allowing for $\mathcal{N}$ to be relatively small, thereby reducing simulation times). The value of $\sigma^2$ can be expressed as

$$\sigma^2 = \text{Var}\left[f(\boldsymbol{Y}) - f(\boldsymbol{X})\right] = \text{Var}\left[f(\boldsymbol{Y})\right] + \text{Var}\left[f(\boldsymbol{X})\right] - 2 \cdot \text{Cov}\left[f(\boldsymbol{Y}), f(\boldsymbol{X})\right]. \qquad (10)$$

---

[3] If $A = \alpha$ in System $X$, and $A = \alpha + \varepsilon$ in System $Y$, then the *forward difference* gives a bias of $\mathcal{O}(\varepsilon)$.

Therefore, if we ensure that $\text{Cov}\left[f(\boldsymbol{Y}), f(\boldsymbol{X})\right]$ is relatively large (compared with $\text{Var}\left[f(\boldsymbol{X})\right]$ and $\text{Var}\left[f(\boldsymbol{Y})\right]$), then $\sigma^2$ will be relatively small. Equivalently, we can ensure that $f(\boldsymbol{X})$ and $f(\boldsymbol{Y})$ are highly correlated[4]. For a constant level of computing resources (i.e., number of simulations, $\mathcal{N}$), a smaller confidence interval will be achieved if we can produce highly correlated samples. The plan is, at each time $t \in [0, T]$, to ensure that the populations of sample paths $\boldsymbol{X}^{(r)}$ and $\boldsymbol{Y}^{(r)}$ are correlated. Then, if $f$ depends on the underlying data, $f\left(\boldsymbol{Y}^{(r)}\right)$ and $f\left(\boldsymbol{X}^{(r)}\right)$ will be correlated, as required.

We now discuss two methods for producing low variance estimators: the Coupled Finite Difference method, described by Anderson[8], and the Common Reaction Path method, described by Rathinam et al.[7]. In the context of this work, we need to clearly distinguish between different finite difference methods that make use of different coupling techniques. As such we will refer to the coupled finite difference method as the Split Propensity Method (SPM) and the common reaction path method as the Common Poisson Process Method (CPM).

## A. Split propensity method

In this section, we discuss the SPM method proposed by Anderson[8] for well-mixed systems. As described in section III, we consider Systems $X$ and $Y$ that differ only in their value for parameter $A$. Thus, for each event $\zeta_j$ in the set of possible events $(\zeta_j)_j$, let $\zeta_j^X$ refer to the event taking place in System $X$, and $\zeta_j^Y$ the same for System $Y$. Furthermore, suppose each $\zeta_j^X$ has propensity $\eta_j^X$, and, likewise, each $\zeta_j^Y$ has propensity $\eta_j^Y$. We will simultaneously simulate pairs of sample paths for Systems $X$ and $Y$, and will enumerate our samples as $[\boldsymbol{X}, \boldsymbol{Y}]^{(r)}$. If $f(\cdot)$ represents the summary statistic of interest, we will use equation (7) to produce partial derivative estimators for equation (6).

---
[4] This follows as $\text{corr}\left[f(\boldsymbol{Y}), f(\boldsymbol{X})\right] = \text{Cov}\left[f(\boldsymbol{Y}), f(\boldsymbol{X})\right] / \left(\text{Var}\left[f(\boldsymbol{X})\right] \text{Var}\left[f(\boldsymbol{Y})\right]\right)$.

---

**Algorithm 3:** The SPM method produces a pair of correlated sample paths: one for System $X$ and another for System $Y$.

---

**Require:** initial conditions, $\boldsymbol{X}(0) = \boldsymbol{Y}(0)$, parameters and terminal time, $T$.
1: set $\boldsymbol{X} \leftarrow \boldsymbol{X}(0)$, and $\boldsymbol{Y} \leftarrow \boldsymbol{Y}(0)$
2: **loop**
3:     calculate propensity values $\eta_j^X$, $\eta_j^Y$ and thus $a_j^C$, $a_j^X$ and $a_j^Y$ (per equations (11))
4:     choose a $a_k^Z$ in proportion to its value, where $k \in \{1, \ldots, J\}$ and $Z \in \{C, X, Y\}$ as per the DM (see algorithm 1)
5:     choose the time to next event, $\Delta$, as per the DM
6:     **if** $t + \Delta > T$ **then**
7:        break
8:     **end if**
9:     **if** $Z \in \{C, X\}$ **then**
10:       set $\boldsymbol{X} \leftarrow \boldsymbol{X} + \kappa_k$
11:     **end if**
12:     **if** $Z \in \{C, Y\}$ **then**
13:       set $\boldsymbol{Y} \leftarrow \boldsymbol{Y} + \kappa_k$
14:     **end if**
15:     set $t \leftarrow t + \Delta$
16: **end loop**

---

The SPM will be implemented with a suitable SSA, such as the DM or MNRM. In order to use algorithm 1 or algorithm 2, we will define a number of *channels*. The *channels* take the role of "events" that are simulated by either the DM or the MNRM. The *channels* will be chosen to ensure that the sample paths for Systems $X$ and $Y$ are highly correlated. Thus, for each *channel* we need to define its propensity (see $\eta_j$ in algorithm 1 or 2), as well the effect of each *channel* on the population levels (see $\kappa_j$ in algorithm 1 or 2). We do this as follows. For each $\zeta_j$, where $j \in \{1, 2, \ldots, J\}$, define the following propensities:

$$a_j^C = \min\left\{\eta_j^X, \eta_j^Y\right\}; \qquad a_j^X = \eta_j^X - a_j^C; \qquad a_j^Y = \eta_j^Y - a_j^C. \tag{11}$$

The set $\left(a_j^Z\right)_{j,Z}$, where $j \in \{1, 2, \ldots, J\}$ and $Z \in \{C, X, Y\}$, provides the propensities for our *channels*. If the *channel* with propensity value $a_k^C$ fires, then events $\zeta_k^X$ and $\zeta_k^Y$ are fired in Systems $X$ and $Y$, respectively. If the channel associated with propensity $a_k^X$ fires, then event $\zeta_k^X$ fires in System $X$, but event $\zeta_j^Y$ does not fire in System $Y$. Similarly, if the channel associated with propensity $a_k^Y$ fires, then System $Y$ is updated with event $\zeta_k^Y$ [5]. An implementation of this simulation algorithm is provided as algorithm 3.

Note that for each $j$, one of $a_j^X$ and $a_j^Y$ will be zero[8]; and the other term should be much smaller than $a_j^C$. When the SPM is run, the channels with propensity of the form $a_j^C$ will typically fire far more often that the channels with propensities $a_j^X$ and $a_j^Y$. This means that, most of the time, events occur simultaneously in Systems $X$ and $Y$, and the sample paths remain close together. In turn, this can lead to a low variance estimator.

## B. Common Poisson process method

In this section, we describe the CPM method described by Rathinam et al.[7]. Consider Systems $X$ and $Y$, where a chosen parameter $A = \alpha$ has been perturbed, and we are estimating the parametric sensitivity given by (7). We can use the Kurtz representation (see equation (3)) to describe the time evolution of a pair of sample paths for Systems $X$ and $Y$.

---

Algorithm 4: This simulates a path $\boldsymbol{X}$, and preserves the firing times of the underlying Poisson processes. This algorithm has been adapted from algorithm 2, and can be used to implement the CPM.

---

**Require:** initial conditions, $\boldsymbol{X}(0)(= \boldsymbol{Y}(0))$ and terminal time, $T$.
1: set $\boldsymbol{X} \leftarrow \boldsymbol{X}(0)$, and set $t \leftarrow 0$
2: for each $\zeta_j$, set $P_j \leftarrow 0$, generate $T_j \leftarrow \text{Exp}(1)$, and store $T_j$ as the first element of list $L_j$
3: **loop**
4:   for each $\zeta_j$, calculate propensity values $\eta_j(\boldsymbol{X}(t))$ and calculate $\Delta_j$ as

$$\Delta_j = \frac{T_j - P_j}{\eta_j}$$

5:   set $\Delta \leftarrow \min_j \Delta_j$, and $k \leftarrow \text{argmin}_j \Delta_j$
6:   **if** $t + \Delta > T$ **then**
7:     **break**
8:   **end if**
9:   set $\boldsymbol{X}(t + \Delta) \leftarrow \boldsymbol{X}(t) + \kappa_k$, set $t \leftarrow t + \Delta$, and for each $\zeta_j$, set $P_j \leftarrow P_j + \eta_j \cdot \Delta$
10:  generate $u \sim \text{Exp}(1)$, then set $T_k \leftarrow T_k + u$ and append $u$ to the end of list $L_k$
11: **end loop**

---

[5] Therefore, our algorithm simulates the times at which each *channel* fires. The firing of a channel leads to an *event* taking place in a system.

We write

$$\boldsymbol{X}(T) = \boldsymbol{X}(0) + \sum_{j=1}^{J} \mathcal{P}_j \left( \int_0^T \eta_j^X(\boldsymbol{X}(t)) \mathrm{d}t \right) \cdot \kappa_j,$$

$$\boldsymbol{Y}(T) = \boldsymbol{Y}(0) + \sum_{j=1}^{J} \mathcal{P}_j \left( \int_0^T \eta_j^Y(\boldsymbol{Y}(t)) \mathrm{d}t \right) \cdot \kappa_j,$$

where the $\mathcal{P}_j$ are unit-rate Poisson processes[6]. The CPM method produces a low variance estimate for equation (7) by using the same set of Poisson processes[7], for sample paths $\boldsymbol{X}$ and $\boldsymbol{Y}$.

This CPM scheme can be implemented by essentially running the MNRM algorithm (see algorithm 2) twice. The procedure is as follows:

- firstly, simulate a sample path for System $X$ with the MNRM. Make sure that the waiting times for each Poisson process are recorded (see algorithm 4);

- secondly, simulate system $Y$ with the MNRM method, but making use of the recorded waiting times (see algorithm 5).

System $X$ is therefore simulated according to the pseudo-code provided in algorithm 4; System $Y$ is then simulated according to algorithm 5. The extension to RDME networks is therefore straightforward[13].

## C. Mathematically representing coupling methods

A useful relationship between the SPM and CPM methods has been rigorously derived by Anderson and Koyama[18]. A mesh, $\pi = \{0 = s_0 < s_1, \ldots s_L = T\}$, is chosen and equation (3) is then restated as

$$\boldsymbol{X}(T) = \boldsymbol{X}(0) + \sum_{j=1}^{J} \sum_{\ell=0}^{L-1} \mathcal{P}_{j\ell} \left( \int_{s_\ell}^{s_{\ell+1}} \eta_j^X(\boldsymbol{X}(t)) \mathrm{d}t \right) \cdot \kappa_j, \qquad (12)$$

where $\mathcal{P}_{j\ell}$ are independent, unit-rate Poisson processes. If we take $L = 1$, then the CPM is recovered; if the mesh is uniformly spaced, then as $L \to \infty$, the SPM is recovered[18].

---

[6] Recall that a Poisson process $\mathcal{P}(t)$ counts the number of arrivals that occur over the time interval $(0, t]$, where the time between arrivals is exponentially distributed, with parameter 1. We can therefore identify a Poisson process with an ordered list of Exp(1) random variables.

[7] Equivalently, the same set of Exp(1) waiting times.

---

Algorithm 5: This simulates a path $\boldsymbol{Y}$, using the waiting times previously generated when simulating a path $\boldsymbol{X}$.

---

**Require:** initial conditions, $\boldsymbol{Y}(0) = \boldsymbol{X}(0)$, terminal time, $T$, and lists $L_k$
 1: set $\boldsymbol{Y} \leftarrow \boldsymbol{Y}(0)$, and set $t \leftarrow 0$
 2: for each $\zeta_j$, set $P_j \leftarrow 0$
 3: for each $\zeta_j$, set $T_j$ to be the first element of list $L_j$, then delete the first element of $L_j$
 4: **loop**
 5:   for each $\zeta_j$, calculate propensity values $\eta_j(\boldsymbol{Y}(t))$ and calculate $\Delta_j$ as

$$\Delta_j = \frac{T_j - P_j}{\eta_j} \quad (13)$$

 6:   set $\Delta \leftarrow \min_j \Delta_j$, and $k \leftarrow \operatorname{argmin}_j \Delta_j$
 7:   **if** $t + \Delta > T$ **then**
 8:     **break**
 9:   **end if**
10:   set $\boldsymbol{Y}(t + \Delta) \leftarrow \boldsymbol{Y}(t) + \kappa_k$, set $t \leftarrow t + \Delta$, and for each $\zeta_j$, set $P_j \leftarrow P_j + \eta_j \cdot \Delta$
11:   **if** $L_k \neq \emptyset$ **then**
12:     let $u$ be the first element of $L_j$: set $T_k \leftarrow T_k + u$, and then delete $u$
13:   **else**
14:     generate $u \sim \operatorname{Exp}(1)$, then set $T_k \leftarrow T_k + u$
15:   **end if**
16: **end loop**

---

### D. Comparing simulation methods

To illustrate the SPM and CPM simulation methods, we begin with an elementary example. The volume $\Omega$ is partitioned into $K = 101$ equally-sized voxels. A single particle is placed in the middle voxel, and is allowed to diffuse. Zero-flux boundary conditions are implemented. We will study the effect of perturbing the diffusion coefficient of this particle. As before, we make two copies of the model, which we label as Systems $X$ and $Y$. The diffusion coefficient for System $X$ is given by $d_x = 0.9$, and for System $Y$ by $d_y = 1.0$.

In order to apply the SPM and CPM methods, we need to focus on a summary statistic of the form (5). We estimate $\mathbb{E}[f(\cdot)]$, where $f(\cdot)$ provides the voxel that contains the particle, at times $t \in \{1, \ldots, 1000\}$. As $\mathbb{E}[f(\boldsymbol{X})] = \mathbb{E}[f(\boldsymbol{Y})]$, the true value of $\mathbb{E}[f(\boldsymbol{Y}) - f(\boldsymbol{X})]$ will be zero. We will now compare and contrast the SPM and CPM simulation methods.

*Split propensity method*

Initially, our model can be described as[8] $X^{51} = Y^{51} = 1$. Thus, the particle can either jump to voxel $V^{52}$ (on the right-hand side of $V^{51}$) or to $V^{50}$ (on the left-hand side). This means that only the events $S^{51} \to S^{52}$ and $S^{51} \to S^{50}$ have non-zero propensities, and therefore the propensities for the combined process envisaged by the SPM (see equations (11)), are given by

$$a^C = 0.9, \quad a^X = 0.0, \quad a^Y = 0.1.$$

There are therefore two possibilities for the first simulation event:

1. diffusion occurs in Systems $X$ and $Y$ (as $a^C = 0.9$ for both $S^{51} \to S^{52}$ and $S^{51} \to S^{50}$);

2. diffusion occurs only in System $Y$ (as $a^Y = 0.1$ for both $S^{51} \to S^{52}$ and $S^{51} \to S^{50}$).

Note that it is impossible, with this construction, for a diffusion event to occur only in System $X$ (as $a^X = 0$ for both $S^{51} \to S^{52}$ and $S^{51} \to S^{50}$). The propensity values show that possibility (2) occurs with probability 0.1. This could happen if, for example, the particle diffuses to the right in System $Y$ only. Then, the dynamics of System $X$ and $Y$ are no longer coupled, as $a^C$ is 0 for all possible events $\zeta_j \in \{1, \ldots, J\}$. This decoupling continues until, after a random period of time, the particles of System $X$ and $Y$ occupy the same voxel.

*Common Poisson process method*

We now implement the CPM method. Consider system $X$. At the first simulation step, the particle can either diffuse to voxel $V^{52}$ (with event $S^{51} \to S^{52}$ taking place) or to $V^{50}$ (when event $S^{51} \to S^{50}$ takes place). According to algorithm 2, the time to the next event is calculated using equation 13, which gives $\Delta^X = \min\{\Delta^X_{S^{51} \to S^{52}}, \Delta^X_{S^{51} \to S^{50}}\}$, as the propensity values of all other events are zero. The appropriate event, $j$, is implemented at time $\Delta^X$, and this process is repeated as per algorithm 2 until the terminal time. Now consider System $Y$. At the first simulation step, the time to the next event is given by: $\Delta^Y = \min\{\Delta^Y_{S^{51} \to S^{52}}, \Delta^Y_{S^{51} \to S^{50}}\}$. The equations that describe the time to the next reaction for Systems $X$ and $Y$ are linearly scaled versions of one another. Thus, at each step, $\Delta^Y$ will take on a different value to $\Delta^X$, but the choice of event, $j$, will be the same. The CPM therefore ensures that events occur in the same order in System $X$ as in System $Y$ (that is, the same order of the particle moving left, right, left, left, and so on), but at different times.

---

[8] Strictly speaking, if we are to follow the notation of section 2, we should write $X^{51}_1$ and $Y^{51}_1$, to indicate that the particle is of species $S_1$, but as there is only one species, we suppress the subscript.

*Comparing numerical performance*

We are now in a position to compare the performance of the SPM and CPM methods. In figure 1 we estimate the value of $\mathbb{E}[f(\boldsymbol{Y}) - f(\boldsymbol{X})]$. In this case, the CPM ensures that the sample values of $[f(\boldsymbol{Y}) - f(\boldsymbol{X})]$ are typically much closer to zero, the expected value, and are therefore more tightly coupled, than the simulations produced with the SPM method. One might think of System $X$ as a slower version of System $Y$. As the CPM method uses the same randomness to simulate the $n$-th event in both Systems $X$ and $Y$ (even if that event takes place at different times in Systems $X$ and $Y$), it performs better than the SPM. In the discussion (section VII A) we explain how the CPM is better at coupling samples paths than the SPM due to the different natural time-scales of Systems $X$ and $Y$.

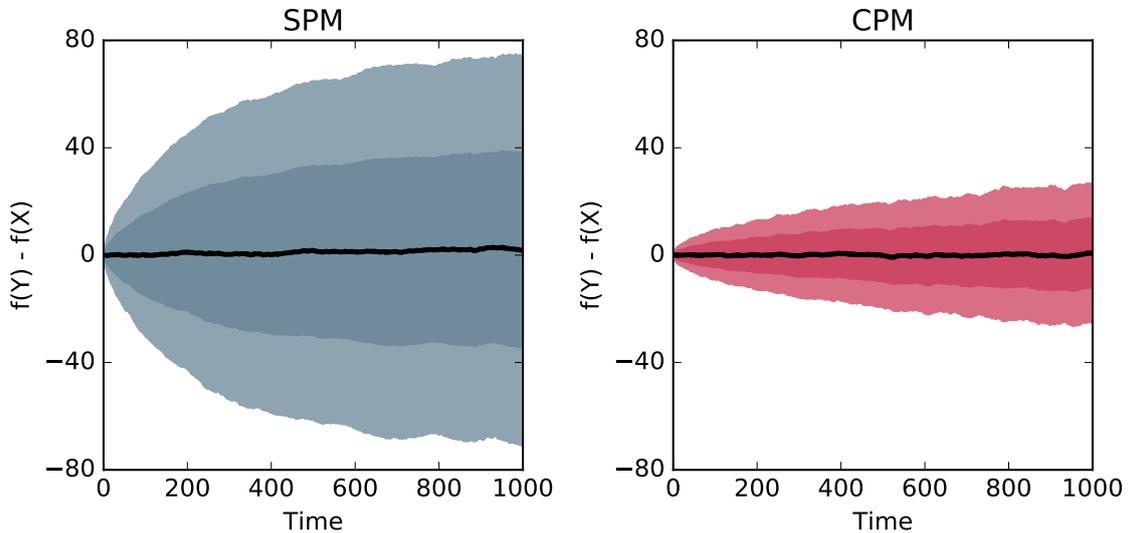

FIG. 1: We compare the SPM (left) and CPM (right) schemes. Each diagram shows the mean value of $f(\boldsymbol{Y}) - f(\boldsymbol{X})$ at different times in black, one standard deviation from the mean in dark shading and two standard deviations in light shading. The estimator is as described in the text and $\mathcal{N} = 10,000$ simulations have been used to produce each plot. The CPM method has a substantially lower variance at each time point.

## IV. COMPARING THE CPM AND SPM METHODS

### A. Case study I

In this first example, we simulate a system which comprises multiple particles. Suppose $\Omega$, of dimensions $L \times a \times a$, is partitioned into $K = 101$ equally-sized voxels along the first dimension. The domain contains particles of species $S_1$ and $S_2$. Suppose particles of species $S_1$ each diffuse with rate $d = D \cdot K^2$, and react to produce particles of type $S_2$ in the following way:

$$S_1 + S_1 \xrightarrow{r} S_1 + S_1 + S_2. \tag{14}$$

We will take $d = 1$ and $r = 0.01$.

As an initial condition, let the central voxel have population $X_1^{51} = 250$, with all other voxels empty. In this example, the summary statistic we are looking to study is the expected total population of species $S_2$ at a terminal time $T$, which is given by summing the $S_2$ populations within each voxel. Thus take $f$ to be

$$\sum_{k=1}^{K} X_2^k(T), \tag{15}$$

so that $\mathbb{Q} = \mathbb{E}\left[f(\boldsymbol{X}(T))\right]$ and we choose $T = 50$. We will evaluate the parametric sensitivities

$$\left.\frac{\partial \mathbb{Q}}{\partial r}\right|_{r=0.01} \quad \text{and} \quad \left.\frac{\partial \mathbb{Q}}{\partial d}\right|_{d=1}, \tag{16}$$

with a suitable confidence interval length.

The finite difference scheme of equation (7) is implemented to estimate the partial derivatives described by equation (16). The numerical simulations can be simplified by not explicitly modelling the diffusion of the $S_2$ particles, because neither the summary statistic given in equation (15), nor reaction (14) is affected by the diffusion of $S_2$. We need to choose the value of the simulation parameter, $\varepsilon$. We start by estimating $\partial \mathbb{Q}/\partial r$. Table I confirms that the SPM and CPM methods can both be used to accurately estimate $\partial \mathbb{Q}/\partial r$ and for each choice of $\varepsilon$ the SPM and CPM methods produce simulations with roughly equal sample variances. The SPM and CPM methods therefore require similar numbers of sample paths to produce estimates with the same confidence interval size.

There are two further points to note. Firstly, as $\varepsilon \downarrow 0$ the sample variance of $[f(\boldsymbol{Y}) - f(\boldsymbol{X})]$, $\sigma^2$, tends to zero. As we are estimating a partial derivative according to (7), the size of the confidence interval given by equation (9) scales as $\sigma/\varepsilon$. Table I shows that $\sigma$ decreases at a slower rate than $\varepsilon$, and consequently more simulations are required to produce estimates with a given confidence interval for smaller choices of $\varepsilon$. The second point we make is that the CPM and SPM methods are substantially more efficient than an uncoupled method. If $f(\boldsymbol{X})$ and $f(\boldsymbol{Y})$ were to be estimated independently, then by equation (10), we have $\sigma^2 \approx 8500$. If $\varepsilon = 2.50 \times 10^{-4}$, then the CPM and SPM each require approximately 35,000 times fewer sample paths than an uncoupled method would require for the same level of statistical accuracy.

We now consider the partial derivative $\partial \mathbb{Q}/\partial d$ evaluated at $d = 1$. Again, the finite difference scheme of equation (7) is implemented, and table II shows a range of estimates for $\partial \mathbb{Q}/\partial d$. As before, a range of choices for $\varepsilon$ are tested. In this case, the CPM method produces a substantially lower estimator variance than the SPM method, and should therefore preferably be used. The relative benefits of the CPM over the SPM are most noticeable when low bias estimates are required (equivalently, when $\varepsilon$ is small). With $\varepsilon = 2.50 \times 10^{-2}$ the CPM method is 6.2 times more efficient (in terms of estimator variance) than the SPM, but when $\varepsilon = 10.00 \times 10^{-2}$, the CPM method is only 2.6 times as efficient as the SPM method. The CPM is therefore particularly useful when low bias estimates for $\partial \mathbb{Q}/\partial d$ are required. Again, both the SPM and CPM are significantly more efficient than an uncoupled method.

In section VII we discuss the differences between the CPM and SPM, and consider intuitive reasons as to why the CPM provides better performance.

### B. Case study II

The second case study considers a stochastic model of the Fisher-KPP wave, which has been used to model the spread of a biological species[19]. We divide a volume $L \times a \times a$ into into $K = 101$ equally-sized voxels along the first dimension. The particles, which are all of the same species, diffuse at a rate $d$ throughout the domain. Within a voxel, the particles interact through the following two reaction channels:

$$R_1 : S \xrightarrow{r_1} S + S; \qquad R_2 : S + S \xrightarrow{r_{-1}} S. \qquad (17)$$

|   | Parameter $\varepsilon$ | Sensitivity estimate | Mean of $[f(\boldsymbol{Y}) - f(\boldsymbol{X})]$ | Variance of $[f(\boldsymbol{Y}) - f(\boldsymbol{X})]$ | Simulations required |
|---|---|---|---|---|---|
| SPM | $2.50 \times 10^{-4}$ | $175{,}882 \pm 502$ | 43.97 | 45.46 | 11098 |
|     | $5.00 \times 10^{-4}$ | $175{,}709 \pm 496$ | 87.85 | 92.54 | 5774 |
|     | $7.50 \times 10^{-4}$ | $175{,}844 \pm 486$ | 131.88 | 139.97 | 4053 |
|     | $10.00 \times 10^{-4}$ | $175{,}509 \pm 501$ | 175.51 | 199.28 | 3055 |
| CPM | $2.50 \times 10^{-4}$ | $175{,}754 \pm 500$ | 43.94 | 46.43 | 11422 |
|     | $5.00 \times 10^{-4}$ | $175{,}586 \pm 498$ | 87.79 | 91.44 | 5675 |
|     | $7.50 \times 10^{-4}$ | $175{,}052 \pm 498$ | 131.29 | 141.06 | 3888 |
|     | $10.00 \times 10^{-4}$ | $175{,}321 \pm 498$ | 175.32 | 189.83 | 2944 |

TABLE I: Estimated values for $\partial \mathbb{Q}/\partial r$ at $r = 0.01$, estimated using equation (7). We have aimed to produce a confidence interval of semi-length 500. The sensitivities appear large: this is because we are working with dimensional quantities.

In order to study this system, we place $10^4$ particles in the left-most voxel (formally, $X^1 = 10^4$), with the remaining voxels left empty. We take $d = 0.1$, $r_1 = 1$ and $r_{-1} = 0.01$; and generate paths until time $T = 25$. As the diffusion rate, $d$, is non-zero, the particles will eventually be able to colonise the whole domain. We focus on two summary statistics of interest:

1. the expected total number of particles in the system at time $T = 25$. This is given by

$$\mathbb{Q}_1 = \mathbb{E}\left[\sum_{k=1}^{K} X^k\right]; \tag{18}$$

|   | Parameter $\varepsilon$ | Sensitivity estimate | Mean of $[f(\boldsymbol{Y}) - f(\boldsymbol{X})]$ | Variance of $[f(\boldsymbol{Y}) - f(\boldsymbol{X})]$ | Simulations required |
|---|---|---|---|---|---|
| SPM | $2.50 \times 10^{-2}$ | $-874.67 \pm 10.00$ | -21.87 | 265.95 | 16342 |
|     | $5.00 \times 10^{-2}$ | $-871.86 \pm 9.90$ | -43.59 | 369.09 | 5783 |
|     | $7.50 \times 10^{-2}$ | $-877.42 \pm 10.10$ | -65.81 | 492.56 | 3296 |
|     | $10.00 \times 10^{-2}$ | $-888.00 \pm 10.06$ | -88.80 | 605.77 | 2298 |
| CPM | $2.50 \times 10^{-2}$ | $-874.97 \pm 10.14$ | -21.87 | 43.76 | 2618 |
|     | $5.00 \times 10^{-2}$ | $-885.38 \pm 10.05$ | -44.27 | 97.20 | 1479 |
|     | $7.50 \times 10^{-2}$ | $-877.60 \pm 9.78$ | -65.82 | 150.38 | 1073 |
|     | $10.00 \times 10^{-2}$ | $-881.90 \pm 9.57$ | -88.19 | 210.37 | 882 |

TABLE II: Estimated values for $\partial \mathbb{Q}/\partial d$ at $d = 1$, estimated using equation (7). We have aimed to produce a confidence interval of semi-length 10.00. The sensitivities appear large: this is because we are working with dimensional quantities.

2. the expected total number of voxels colonized by the population at time $T = 25$. This is evaluated as

$$\mathbb{Q}_2 = \mathbb{E}\left[\sum_{k=1}^{K} \mathbb{I}_{\{X^k > 0\}}\right]. \tag{19}$$

Suppose that the diffusion rate, $d$, is perturbed. As before, the sensitivity of summary statistics given by equations (18) and (19), with respect to a changing diffusion constant, is to be estimated by equation (7). In table III we show estimated values for $\partial \mathbb{Q}_1/\partial d$ (as per equation (18)) and in table IV we show estimated values for $\partial \mathbb{Q}_2/\partial d$ (as per equation (19)). In both cases, the CPM outperforms the SPM.

In this section, we have shown that the optimal finite difference method depends on the model of interest. This is a departure from previous experience, which suggested that the SPM should be preferred[8]. We now discuss two novel simulation strategies.

|  | Parameter $\varepsilon$ | Sensitivity estimate | Mean of $[f(\boldsymbol{Y}) - f(\boldsymbol{X})]$ | Variance of $[f(\boldsymbol{Y}) - f(\boldsymbol{X})]$ | Simulations required |
|---|---|---|---|---|---|
| SPM | $2.50 \times 10^{-3}$ | $6,230 \pm 251$ | 15.57 | 2,512.10 | 24464 |
| SPM | $5.00 \times 10^{-3}$ | $6,406 \pm 251$ | 32.03 | 4,809.75 | 11715 |
| SPM | $7.50 \times 10^{-3}$ | $6,348 \pm 245$ | 47.61 | 6,145.21 | 6991 |
| SPM | $10.00 \times 10^{-3}$ | $6,220 \pm 250$ | 62.20 | 8,478.50 | 5195 |
| CPM | $2.50 \times 10^{-3}$ | $6,303 \pm 249$ | 15.76 | 591.91 | 5878 |
| CPM | $5.00 \times 10^{-3}$ | $6,509 \pm 252$ | 32.54 | 1,156.78 | 2800 |
| CPM | $7.50 \times 10^{-3}$ | $6,317 \pm 250$ | 47.38 | 1,571.85 | 1721 |
| CPM | $10.00 \times 10^{-3}$ | $6,266 \pm 247$ | 62.66 | 1,868.66 | 1176 |

TABLE III: Estimated values for $\partial \mathbb{Q}_1/\partial d$ at $d = 1$, estimated using (7). We have aimed to produce a confidence interval of semi-length 250. The sensitivities appear large: note this is because we are working with dimensional quantities.

|   | Parameter $\varepsilon$ | Sensitivity estimate | Mean of $[f(\boldsymbol{Y}) - f(\boldsymbol{X})]$ | Variance of $[f(\boldsymbol{Y}) - f(\boldsymbol{X})]$ | Simulations required |
|---|---|---|---|---|---|
| SPM | $2.50 \times 10^{-3}$ | $71.83 \pm 2.51$ | $17.96 \times 10^{-2}$ | $40.17 \times 10^{-2}$ | 39140 |
| SPM | $5.00 \times 10^{-3}$ | $72.40 \pm 2.51$ | $36.20 \times 10^{-2}$ | $75.05 \times 10^{-2}$ | 18378 |
| SPM | $7.50 \times 10^{-3}$ | $72.63 \pm 2.48$ | $54.47 \times 10^{-2}$ | $98.12 \times 10^{-2}$ | 10936 |
| SPM | $10.00 \times 10^{-3}$ | $72.67 \pm 2.50$ | $72.67 \times 10^{-2}$ | $130.20 \times 10^{-2}$ | 7977 |
| CPM | $2.50 \times 10^{-3}$ | $74.34 \pm 2.51$ | $18.58 \times 10^{-2}$ | $16.44 \times 10^{-2}$ | 16051 |
| CPM | $5.00 \times 10^{-3}$ | $71.67 \pm 2.50$ | $35.84 \times 10^{-2}$ | $26.82 \times 10^{-2}$ | 6594 |
| CPM | $7.50 \times 10^{-3}$ | $71.25 \pm 2.49$ | $53.44 \times 10^{-2}$ | $32.86 \times 10^{-2}$ | 3634 |
| CPM | $10.00 \times 10^{-3}$ | $74.22 \pm 2.51$ | $74.22 \times 10^{-2}$ | $38.10 \times 10^{-2}$ | 2320 |

TABLE IV: Estimated values for $\partial \mathbb{Q}_2/\partial d$ at $d = 1$, estimated using (7). We have aimed to produce a confidence interval of semi-length 250. The sensitivities appear large: this is because we are working with dimensional quantities.

## V. GROUPED SAMPLING

This new approach reduces the sample variance of equation (7) for spatially-extended reaction networks. Consider again the SPM method, where we have enumerated the events that change the population matrix, equation (1), as $(\zeta_j)_j$. The propensity values of two systems, labeled as systems $X$ and $Y$, are inserted into equations (11), and a sample path for each system is then generated. We argued by example in section III D that the values given by equation (11) are very sensitive to the exact location of particles, and so it can be difficult to generate tightly coupled sample paths. The Grouped Sampling Method (GSM) is designed to be less sensitive to the exact configuration of each of Systems $X$ and $Y$, and can therefore achieve a lower sample variance under a wider variety of circumstances. We explain the GSM by first considering the simulation of a single system. We can partition the set of events $(\zeta_j)_j$ that change the population matrix into $(M + 2)$ groups. The groups are as follows: $\Gamma_1$ contains all $R_1$ reaction events (so that there is an entry for each voxel, meaning $K$ events are contained in $\Gamma_1$), $\Gamma_2$ (which contains all $R_2$ events), ..., $\Gamma_M$ contains all $R_M$ reaction events, $\Gamma_{M+1}$ contains all diffusive jumps where particles diffuse to the left, and, $\Gamma_{M+2}$ contains events where particles diffuse to the right. Furthermore, we order the events inside each group according to the voxel in which they take place (the importance of this will soon become clear). For example, an event that takes place in voxel $V^3$ will be "next to" an event that takes place in voxel $V^4$. We can therefore simulate the events which take place in a single sample path by:

1. randomly select a group, $\Gamma_g$, where the probability of group $\Gamma_g$ being chosen is proportional to the sum of the propensities of the events inside that group;

2. randomly choose an event $\zeta_k$ in group $\Gamma_g$, where the probability of $\zeta_k$ being chosen is proportional to its propensity value.

This approach clearly produces the same dynamics as algorithms 1 or 2. We now show how to use this two-step method to simulate a correlated pair of sample paths. If we follow the

---

Algorithm 6: The grouped sampling method is a two-step simulation method. We share randomness between Systems $X$ and $Y$ at each of the two steps. A group $\Gamma_g$ is chosen according to the SPM; the SPM also decides whether an event takes place in both Systems $X$ and $Y$, or only one. At the second step, the inverse transform method chooses the voxel wherein the reactants are found.

---

**Require:** initial conditions, $\boldsymbol{X}(0) = \boldsymbol{Y}(0)$, terminal time, $T$, and event groups $(\Gamma_g)$.
1: set $\boldsymbol{X} \leftarrow \boldsymbol{X}(0)$, $\boldsymbol{Y} \leftarrow \boldsymbol{Y}(0)$ and $t \leftarrow 0$
2: **loop**
3:    for each $\Gamma_g$, calculate $a_g^C$, $a_g^X$ and $a_g^Y$ according to equations (20)
4:    set $a_0 \leftarrow \sum_g \left(a_g^C + a_g^X + a_g^Y\right)$ and take $\Delta \leftarrow \mathrm{Exp}(a_0)$
5:    **if** $t + \Delta > T$ **then**
6:       break
7:    **end if**
8:    choose $(g, Z)$ with probability $a_g^Z/a_0$, where $Z \in \{C, X, Y\}$.
9:    **if** $Z = C$ **then**
10:       set $u \leftarrow U(0,1)$, and choose $k^X$, $k^Y$ with inverse transform method $\mathrm{inv}(u,g)$ (see main text)
11:       set $\boldsymbol{X} \leftarrow \boldsymbol{X} + \kappa_{k^X}$ and $\boldsymbol{Y} \leftarrow \boldsymbol{Y} + \kappa_{k^Y}$
12:    **else if** $Z = X$ **then**
13:       choose $k$ using $\mathrm{inv}(u,g)$, and set $\boldsymbol{X} \leftarrow \boldsymbol{X} + \kappa_k$
14:    **else if** $Z = Y$ **then**
15:       choose $k$ using $\mathrm{inv}(u,g)$, and set $\boldsymbol{Y} \leftarrow \boldsymbol{Y} + \kappa_k$
16:    **end if**
17:    set $t \leftarrow t + \Delta$
18: **end loop**

two-step procedure, there are two opportunities to share random numbers between Systems $X$ and $Y$. At the first step outlined above, we effectively choose an event type: be it diffusion to the left, to the right, or a reaction of type $R_j$. This step will be accomplished by using the SPM method described in section III A. The SPM method decides whether an event takes place in System $X$, $Y$ or both. At the second step, we choose which voxel the event takes place in. This step will be performed with an inverse transform method[9]. This means that it is possible for the same event (e.g. diffusion to the left) to take place at the same time in both Systems $X$ and $Y$, even if that event takes place in a slightly different voxel (note that if we did not order the events inside each group, this would not be possible). In order to use the SPM to perform step (1) of the simulation, for each group $\Gamma_g$, we define the following propensities

$$a_g^C = \min\left\{\sum_{\zeta_j \in \Gamma_g} \eta_j^X, \sum_{\zeta_j \in \Gamma_g} \eta_j^Y\right\}, \qquad a_g^X = \sum_{\zeta_j \in \Gamma_g} \eta_j^X - a_j^C, \qquad a_g^Y = \sum_{\zeta_j \in \Gamma_g} \eta_j^Y - a_j^C, \quad (20)$$

where $\eta_j^X$ and $\eta_j^Y$ refer to the propensity of event $\zeta_j$ in Systems $X$ and $Y$, respectively.

To perform step (2), we describe an inverse transform method, $\text{inv}(u, g)$, where $u$ is a uniformly generated $(0, 1)$ random variable, and $g$ refers to the index of the ordered group $\Gamma_g$. Pseudo-code is provided in algorithm 6; we implement the algorithm in the next section.

## A. Illustrating grouped sampling

The grouped sampling method is now tested and compared with the ungrouped SPM and CPM. We will return to case study I, where species $S_1$ diffuses through a volume of $K = 101$ voxels, and reacts to form $S_2$ particles. In order to ensure that the simulation results presented in table I and table II are not simply a consequence of the symmetry of case study I, we introduce stochastic drift into the model, so that the diffusion of the $S_1$ particles is biased to either the left or the right. Thus, for each voxel, $V^k$, the $S_1$ particles

---

[9] The inverse transform method proceeds as follows: take $u \in (0, 1)$ as a uniformly-distributed random input. Write down the ordered events inside $\Gamma_g$ as $\zeta_1, \zeta_2, \ldots, \zeta_J$. Each event $\zeta_j$ has associated propensity $\eta_j$. Then, the inverse transform methods chooses event $j^*$, where

$$\sum_{j=1}^{j^*-1} \eta_j < u \sum_{j=1}^{J} \eta_j < \sum_{j=1}^{j^*} \eta_j.$$

.

diffuse according to

$$S_1^k \xrightarrow{d_\ell} S_1^{k-1}, \qquad S_1^k \xrightarrow{d_r} S_1^{k+1}; \qquad (21)$$

where $d_\ell$ and $d_r$ are the appropriate biased transition rates. As before, we enforce zero-flux conditions. We take $d_\ell = 3.5$, $d_r = 1.0$ (meaning a net drift rate of 2.5), and $T = 50$. We re-use the remaining parameters and initial conditions from case study I. We consider again the expected total number of $S_2$ particles in the system (given by equation (15)), and will now estimate the value of

$$\left.\frac{\partial \mathbb{Q}}{\partial d_\ell}\right|_{d_\ell = 3.5}, \qquad (22)$$

where $\mathbb{Q} = \mathbb{E}[f(\boldsymbol{X})]$. The results of using the finite difference scheme given by equation (16) are given in table V. The results show that the grouped sampling method is substantially more efficient than the SPM method (requiring approximately three times fewer sample paths), and provides a modest improvement over the CPM method (a saving of approximately 30%).

|  | Parameter $\varepsilon$ | Sensitivity estimate | Mean of $[f(\boldsymbol{Y}) - f(\boldsymbol{X})]$ | Variance of $[f(\boldsymbol{Y}) - f(\boldsymbol{X})]$ | Simulations required |
|---|---|---|---|---|---|
| SPM | 0.25 | $4,589 \pm 10$ | 1,147.27 | 6,451.29 | 3962 |
| SPM | 0.50 | $4,598 \pm 10$ | 2,298.83 | 13,269.47 | 2028 |
| SPM | 0.75 | $4,616 \pm 10$ | 3,462.18 | 18,696.09 | 1271 |
| CPM | 0.25 | $4,588 \pm 10$ | 1,146.98 | 3,424.79 | 2078 |
| CPM | 0.50 | $4,593 \pm 10$ | 2,296.26 | 6,737.83 | 1040 |
| CPM | 0.75 | $4,612 \pm 10$ | 3,459.29 | 10,126.12 | 685 |
| GPM | 0.25 | $4,587 \pm 10$ | 1,146.79 | 2,237.95 | 1376 |
| GPM | 0.50 | $4,599 \pm 10$ | 2,299.58 | 4,205.92 | 647 |
| GPM | 0.75 | $4,612 \pm 10$ | 3,458.76 | 7,363.30 | 500 |

TABLE V: Estimated values for $\partial \mathbb{Q}/\partial d_\ell$ at $d_\ell = 3.5$, estimated using equation (7). We have aimed to produce a confidence interval of semi-length 10.

## VI. HYBRID SAMPLING

This is our second development. We have already demonstrated that the relative performance of the CPM and SPM methods depend on the problem to be investigated. In this section, we describe a hybrid switching scheme, which we call the multichoice (MC) method. The benefits of the multichoice method are two-fold: firstly, it lets us choose the best coupling method (either the SPM or CPM) for each event; and, secondly, it allows us to dynamically change this decision. This flexibility enables an even greater reduction in sample variance. Our method has been designed without any specific problem in mind, but a heuristic justification for our approach is included within the discussion (see section VII B).

To improve on the SPM and CPM methods, we will dynamically assign each event $\zeta_j$, $j \in \{1, \ldots, J\}$, to either the "SPM part" or "CPM part" of the algorithm, to determine how it should be simulated. This assignment may change, depending on the population matrix (1) of the system. To simplify matters, we will do this on a voxel-by-voxel basis, so that events with reactants in the same voxel will all be simulated with the SPM, or alternatively with the CPM. We first provide a broad overview of the method, and then explain how it might be implemented. Our method for switching between the SPM and CPM is intricate, because it is constructed in a way that ensures there is no additional bias introduced into the estimation of summary statistics. We return to this point in section VI B.

## A. Switching between the SPM and CPM

We might think of using the voxel populations to decide whether the SPM or CPM should be used for each voxel, $V^k$, and then labeling our decision as $\Phi(X^k, Y^k)$. This suggests that every time the population matrix of System $X$ or $Y$ changes, we check if the new populations lead to a different choice. If necessary, we immediately change coupling methods. In section VI B, we show that this method will result in a further, uncontrolled bias in summary statistics.

We avoid biasing the statistics by implementing the following procedure. Consider only System $X$. For each value of $X^k$ (the population values in voxel $V^k$), we will determine which of the SPM or CPM is likely to be the better coupling method to implement (without explicitly considering $Y^k$). We label this method as $\Psi(X^k)$, and impose the method for voxel $V^k$ in System $X$. When the value of $X^k$ changes, we see if the coupling choice for voxel $V^k$ changes. The same procedure is followed for System $Y$ (with the label $\Psi(Y^k)$). When the populations $X^k$ and $Y^k$ both suggest the same method (either the SPM or CPM be used) for voxel $V^k$, then this method is implemented. However, there can be an interface period where the values of $X^k$ and $Y^k$ mean require different coupling methods to be implemented, and a bespoke simulation approach is needed for this *interface region*. This interface region is required to make sure that no bias is introduced. The upside is that in scenarios we have encountered, the size of the interface region is small relative to the time-scale of the sample path. We explain the details of the multichoice method in two steps. We first describe the multichoice method for just a single voxel. The second step is to implement the multichoice method on a system with many voxels.

### *Considering a single voxel*

Initially, suppose that $\Psi(X) = \Psi(Y) = \text{CPM}$. The CPM is implemented, and therefore:

- algorithm 4 is used in System $X$ and algorithm 5 in System $Y$.

Now suppose that $\Psi(X)$ changes to SPM, but $\Psi(Y)$ remains as the CPM. This is an interface region, which is characterized by System $X$ transitioning from the CPM to the SPM. Thus,

- algorithm 2 is used in System $X$ and algorithm 5 continues to be used in System $Y$.

Note that the lists of arrival times ($L_j$ for $j \in \{1, \ldots, J\}$) generated by algorithm 4 are still used at this step. Next, suppose that $\Psi(Y)$ changes to SPM. This means that we can couple the paths and

- algorithm 3 is used for both Systems $X$ and $Y$.

The lists ($L_j$) for $j \in \{1, \ldots, J\}$ are deleted. Next, suppose $\Psi(X)$ changes to CPM, but $\Psi(Y)$ remains as the SPM. We are again in an interface region, and therefore

- algorithm 4 is used in System $X$, and algorithm 2 in System $Y$.

Finally, suppose that $\Psi(Y)$ changes to CPM. Full coupling of the paths can now be achieved, and:

- algorithm 4 is used in System $X$; algorithm 5 is restarted in System $Y$.

Note that algorithm 5 uses a new lists of arrival times ($L_j$). This scenario is graphically illustrated in figure 2.

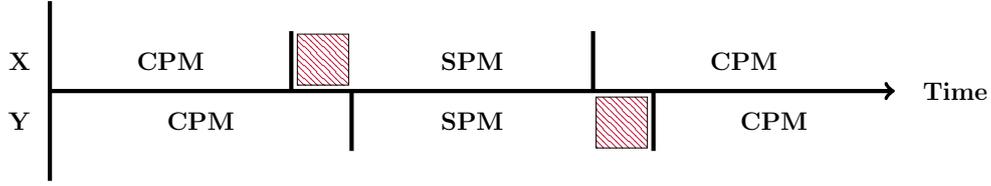

FIG. 2: This diagram illustrates the multichoice coupling method. The hatching illustrates an interface region: please see section VI A above for further information.

The above scenario described two kinds of interface scenarios: (1) a system has moved from the CPM to the SPM, with the other system to follow; and (2) a system has moved from the SPM to the CPM, with the other system to follow. This exhausts all possible ways in which an interface region can arise.

*Considering multiple voxels*

The multichoice method is now implemented across multiple voxels. There are multiple ways of achieving this, but we describe a method that is relatively easy to implement. Our procedure is to first simulate a sample path for System $X$ using algorithm 4. We record the population matrix $\boldsymbol{X}$ at each time $t$, and save the firing times of the Poisson processes, $\mathcal{P}_j$, used to generate the sample path for $X$. Now that sample path for System $X$ has been simulated, we must produce a sample path for System $Y$. For each voxel of System $Y$, there are three possible simulation algorithms:

1. $\Psi(X^k) = \Psi(Y^k) = $ CPM, so that algorithm 5 is implemented;

2. $\Psi(X^k) \neq \Psi(Y^k)$, so either algorithm 5 or algorithm 2 is required;

3. $\Psi(X^k) = \Psi(Y^k) = $ SPM, so that the SPM method must be used.

Recall that at this stage, we have fully simulated the sample path for System $X$. Cases (1) and (2) above are straightforward to deal with. To deal with the third possibility, we will need to reverse-engineer the SPM to deduce the sample path for System $Y$.

The SPM is reverse-engineered as follows: for each event that changes the population levels of System $X$, we need to decide whether it also takes place in System $Y$. At each simulation step, we compare propensities. If $\eta_j^X < \eta_j^Y$, then:

---

Algorithm 7: The SPM simulates path $\boldsymbol{Y}$ by sharing randomness with a previously generated path $\boldsymbol{X}$. This method produces the same dynamics as algorithm 3.

---

**Require:** initial conditions, $\boldsymbol{Y}(0) = \boldsymbol{X}(0)$, terminal time, $T$, and lists $L_k$
1: set $\boldsymbol{Y} \leftarrow \boldsymbol{Y}(0)$, and set $t \leftarrow 0$
2: for each $\zeta_j$, set $P_j^X \leftarrow 0$
3: for each $\zeta_j$, set $T_j^X$ to be the first element of list $L_j$, then delete the first element of $L_j$
4: for each $\zeta_j$, set $P_j^Y \leftarrow 0$, and generate $T_j^Y \leftarrow \text{Exp}(1)$
5: **loop**
6:    for each $\zeta_j$, calculate propensity values $\eta_j^X(\boldsymbol{X}(t))$, $\eta_j^Y(\boldsymbol{Y}(t))$
7:    for each $\zeta_j$ and $Z \in \{X, Y\}$, calculate $\Delta_j^Z$ as

$$\Delta_j^X = \frac{T_j^X - P_j^X}{\eta_j^X}, \qquad \Delta_j^Y = \frac{T_j^Y - P_j^Y}{\max\{0, \eta_j^Y - \eta_j^X\}}$$

8:    set $\Delta \leftarrow \min_{j,Z} \Delta_j^Z$ (where the minimum is over $Z$ and $j$), and let $k \leftarrow \operatorname{argmin}_{j,Z} \Delta_j^Z$
9:    **if** $t + \Delta > T$ **then**
10:      break
11:    **end if**
12:    **if** $Z = X$ **then**
13:      **if** $\eta_j^X \leq \eta_j^Y$ **then**
14:         set $\boldsymbol{Y}(t + \Delta) \leftarrow \boldsymbol{Y}(t) + \kappa_k$
15:      **else if** $\eta_j^X > \eta_j^Y$ **then**
16:         with probability $\eta_j^y/\eta_j^X$, set $\boldsymbol{Y}(t + \Delta) \leftarrow \boldsymbol{Y}(t) + \kappa_k$
17:      **end if**
18:      set $t \leftarrow t + \Delta$
19:      **if** $L_k \neq \emptyset$ **then**
20:         let $u$ be the first element of $L_j$: set $T_k^X \leftarrow T_k^X + u$, and then delete $u$
21:      **else**
22:         generate $u \sim \text{Exp}(1)$, then set $T_k^X \leftarrow T_k + u$
23:      **end if**
24:    **else if** $Z = Y$ **then**
25:      set $\boldsymbol{Y}(t + \Delta) \leftarrow \boldsymbol{Y}(t) + \kappa_k$, and set $t \leftarrow t + \Delta$
26:      generate $u \sim \text{Exp}(1)$, then set $T_k \leftarrow T_k + u$
27:    **end if**
28:    for each $\zeta_j$, set $P_j^X \leftarrow P_j^X + \eta_j^X \cdot \Delta$ and set $P_j^Y \leftarrow P_j^Y + \max\{0, \eta_j^Y - \eta_j^X\} \cdot \Delta$
29: **end loop**

---

- any $\zeta_j$ that takes place in System $X$ necessarily also takes place in System $Y$ (see equation (11), where $a_j^X = 0$); and

- it is possible for an event $\zeta_j$ to fire only in System $Y$ (in terms of equation (11), $a_j^Y \geq 0$).

However, if $\eta_j^X > \eta_j^Y$:

- if an event $\zeta_j$ fires in System $X$, then it fires with probability $\eta_j^Y/\eta_j^X$ in System $Y$ (in terms of equation (11), $a_j^C = \eta_j^Y$, $a_j^X \geq 0$ and $a_j^Y = 0$).

Please see algorithm 7 for a pseudo-code implementation of the SPM.

Finally, we are in a position to describe the overall multichoice method. Each of the aforementioned algorithms are modifications of algorithm 2, and so combining them into a single algorithm is natural. This technique is described in full in algorithm 8.

## B. A warning about model bias

In this section, we briefly explain why deciding on the coupling method based on the current populations $X^k$ and $Y^k$ only leads to a model bias. Consider Systems $X$ and $Y$. For each voxel $V^k$, we might think of using the voxel populations to decide whether the SPM or CPM should be used for that voxel, and then labeling our decision as $\Phi(X^k, Y^k)$. Every

---

Algorithm 8: The multi-choice algorithm simulates path $Y$ by using randomness from path $\boldsymbol{X}$. Different simulation methods are used for different voxels.

---

**Require:** initial conditions, $\boldsymbol{Y}(0) = \boldsymbol{X}(0)$, terminal time, $T$, and complete details of $\boldsymbol{X}$
1: set $\boldsymbol{Y} \leftarrow \boldsymbol{Y}(0)$, and set $t \leftarrow 0$
2: for each $V^k$, let $M_k$ be simulation method implied by $\Psi(X^k)$ and $\Psi(Y^k)$;
3: **for** each voxel $V^k$ **do**
4:     configure $P_j$, $T_j$, etc. as appropriate per $M_k$
5: **end for**
6: **loop**
7:     **for** each voxel $V^k$ **do**
8:        calculate propensities, required internal values, set $\underline{\Delta}^k$ to be time to next event
9:     **end for**
10:     **if** $t + \min \underline{\Delta}^k > T$ **then**
11:        break
12:     **end if**
13:     set $t \leftarrow t + \min \underline{\Delta}^k$ and update $\boldsymbol{Y}$ per $\arg\min \underline{\Delta}^k$
14:     **for** each voxel $V^k$ **do**
15:        perform housekeeping as required by $M_k$
16:        recalculate $\Psi(X^k)$, $\Psi(Y^k)$ and so $M_k$ at time $t$
17:     **end for**
18: **end loop**

---

time one of $X^k$ or $Y^k$ changes, $\Phi$ is re-evaluated. When our choice $\Phi$ changes, we might then hope to immediately change the coupling method by relying on the memoryless property of exponential variates. Unfortunately, this implementation we have described leads to a model bias. Suppose that at time $t = 0$, $\Phi(X^k, Y^k) = \text{CPM}$, and so the events taking place in $V^k$ are simulated by explicitly considering the arrival times of Poisson processes (recall the System $X$ and $Y$ share Poisson processes). If an event fires at time $t = t^*$ that results in $\Phi(X^k, Y^k) = \text{SPM}$, we immediately switch to the SPM method. Over the time interval $(0, t^*]$ the Poisson processes associated with voxel $V^k$ might have fired a different number of times in Systems $X$ and $Y$. Let us suppose, without loss of generality, that the Poisson process $\mathcal{P}_j$ fires more times in System $X$ than in System $Y$. By immediately switching to the SPM method, we stop using the CPM method, and so the firings of $\mathcal{P}_j$ that have been ear-marked to occur in System $Y$, do not take place. The difficulty is that these ear-marked arrivals have already affected the value of $X^k$, thereby contributing to the choice $\Phi(X^k, Y^k) = \text{SPM}$. As the ear-marked values play a role in changing $\Phi$, when we observe the change in $\Phi$ we gain information as to the distribution of the ear-marked arrival times, and can no longer assume that they are exponentially distributed. We therefore cannot use the memoryless property on these arrival times without introducing a bias. The multichoice method will not bias model statistics for the following reason: when an individual system changes coupling methods (from the CPM to SPM, for example), this is done on the basis of the random numbers that have already been simulated and used in producing that sample path. The random numbers which that be simulated in future, have no role in the coupling method changing, and we can therefore safely discard them.

|  | Parameter $\varepsilon$ | Sensitivity estimate | Mean of $[f(\boldsymbol{Y}) - f(\boldsymbol{X})]$ | Variance of $[f(\boldsymbol{Y}) - f(\boldsymbol{X})]$ | Simulations required |
|---|---|---|---|---|---|
| $\partial \mathbb{Q}_1 / \partial d$ | $2.50 \times 10^{-3}$ | $6,446 \pm 250$ | 16.12 | 107.22 | 1057 |
|  | $5.00 \times 10^{-3}$ | $6,657 \pm 252$ | 33.28 | 215.90 | 523 |
|  | $7.50 \times 10^{-3}$ | $6,572 \pm 260$ | 49.29 | 365.00 | 370 |
|  | $10.00 \times 10^{-3}$ | $6,066 \pm 241$ | 60.66 | 469.35 | 310 |
| $\partial \mathbb{Q}_2 / \partial d$ | $2.50 \times 10^{-3}$ | $73.40 \pm 2.53$ | $18.35 \times 10^{-2}$ | $16.63 \times 10^{-2}$ | 16000 |
|  | $5.00 \times 10^{-3}$ | $71.23 \pm 2.50$ | $35.62 \times 10^{-2}$ | $26.86 \times 10^{-2}$ | 6612 |
|  | $7.50 \times 10^{-3}$ | $72.30 \pm 2.52$ | $54.22 \times 10^{-2}$ | $34.84 \times 10^{-2}$ | 3753 |
|  | $10.00 \times 10^{-3}$ | $72.14 \pm 2.52$ | $72.14 \times 10^{-2}$ | $39.96 \times 10^{-2}$ | 2416 |

TABLE VI: Estimated values for $\partial \mathbb{Q}_1 / \partial d$ and $\partial \mathbb{Q}_2 / \partial d$ at $d = 1.0$, estimated using (7) and the multichoice method. Appropriate confidence intervals have been constructed.

C. Case study II

We return to case study II, which concerns a stochastic Fisher-KPP wave. There are two distinct behaviour to consider. Between the wave-front and the left boundary, high molecular populations are maintained. At the wave-front, diffusion drives the wave to the right. The colonisation of the domain is due to a small number of molecules jumping to the right. We postulate that the CPM method might work better for simulating events when molecular populations are low, and that the SPM method should be preferred in the case of high molecular numbers that are maintained at a steady state. Thus, we summarise our choice of coupling method, $\Psi$, as

$$\Psi(X^k) = \begin{cases} \text{CPM}, & \text{if } X^k \leq \alpha; \\ \text{SPM}, & \text{if } X^k > \alpha. \end{cases}$$

where $\alpha$ is a chosen threshold. We have worked with $\alpha = 67$, and will use this throughout the rest of this section. We have chosen $\alpha$ to be well away from the favourable state, but equally, not so low so that the benefits of the CPM cannot be realised. Further information as to the heuristics of choosing a coupling method are provided in the discussion. In our experience, the algorithm is not particularly sensitive to the precise choice of $\alpha$.

The multichoice method is now implemented. The summary statistics of interest are the total number of particles (see (18)) and the number of non-empty voxels (see (19)). In table VI we set out the results of our investigation into the partial derivatives given by (18) and (19) with respect to a change in the diffusion rate, $d$. We compare the simulation results in table III and table IV. In the case of the sensitivity of the total number of particles (see (18)), we see that the multichoice method can be up to 5.6 times more efficient as the CPM method, and 23 times more efficient as the SPM method. When considering the sensitivity of the total number of voxels occupied (see (19)), we see that, as expected, the multichoice method provides roughly equivalent performance. These speed-ups are shown in figure 3.

## VII. DISCUSSION

In this work, we have shown that the SPM and CPM methods for estimating parametric sensitivities in well-mixed systems can be naturally extended to study spatially-inhomogeneous RDME models. Previous work proceeded on the assumption that the SPM provides lower-variance estimates than the CPM method, and should therefore be preferred[8]. We have shown that the relative performance of each method depends on the model of interest, as well as the summary statistics that are to be computed. In addition, we have presented two new simulation strategies: firstly, a grouped sampling method; and, secondly, a hybrid method that dynamically combines the SPM and CPM approaches. The efficiency of these novel methods have been demonstrated with numerical examples.

In the remainder of this work, we discuss a number of unresolved issues and challenges. We provide some intuition as to the circumstances under which the CPM method outperforms the SPM technique, and when grouped sampling or the multichoice method are required. We then discuss a number of implementation issues.

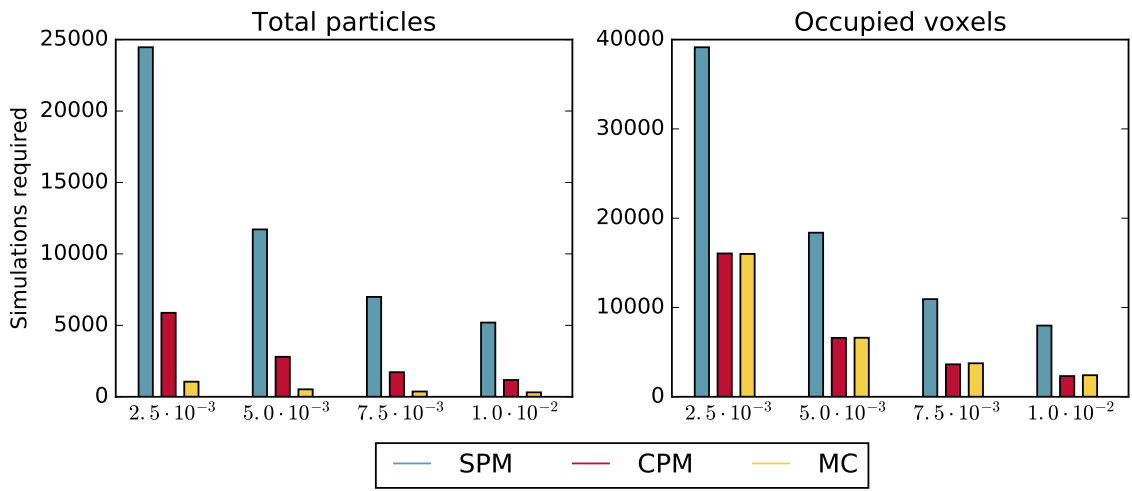

FIG. 3: This shows the number of simulations required to estimate partial derivatives for the Fisher-KPP system, $\partial \mathbb{Q}_1/\partial d$ and $\partial \mathbb{Q}_2/\partial d$ at $d = 1.0$, for a constant level of statistical accuracy. A range of values for $\varepsilon$ are shown on the $x$-axes. For further information, see table III, table IV, and table VI.

## A. Intuitive differences between the SPM and CPM methods

Suppose that we are producing sample paths for Systems $X$ and $Y$ over a time interval $[0, T]$. Informally, the SPM compares the propensities of each event, $\zeta_j$, in Systems $X$ and $Y$. If the propensities for event $\zeta_j$ are exactly the same, then we can insist that whenever $\zeta_j$ takes place in one of Systems $X$ or $Y$, it also takes place in the other system. If the propensities are different, then if $\zeta_j$ takes place in one system, it also takes place in the other system with some probability. If the propensities are similar, then the aforementioned probability will be high, and so we expect that the processes will be tightly coupled. This procedure is Markovian in the sense that it only depends on the current propensity values.

The CPM method is different. For each event, $\zeta_j$, a single, unit-rate Poisson process is used to simulate events for both Systems $X$ and $Y$. We determine the firing times of event $\zeta_j$ by keeping track of internal times (see equation (3)), and comparing them with the firing times of each unit-rate Poisson process, $\mathcal{P}_j$. The internal times depend on the entire history of the sample path, and not only the present value. The sequence of arrival times for $\mathcal{P}_j$ is kept the same, and the CPM coupling method therefore uses the same arrival time for the $n$-th firing of the Poisson process. Unlike the SPM, this coupling is not explicitly time-based.

Sometimes the SPM and CPM methods produce estimates for equation (7) that have similar variances, as seen in case study I with a perturbed reaction rate, $r$. In cases where a perturbed parameter means one process has a different natural time-scale to the other, then the CPM method provides better performance. In case study I, when the diffusion rate, $d$, was perturbed, System $X$ and $Y$ operated on different natural time-scales, with System $Y$ effectively a faster version of System $X$. A time-based coupling provides inferior performance. There are conditions under which the SPM method is likely to outperform the CPM technique. In particular, where a steady state is expected, the SPM coupling is memoryless, which allows for mean-reversion effects. When molecular populations are high, the SPM method might be expected to perform better than it would when molecular populations are low. The summary statistic of interest will also have an effect on the choice between the SPM and CPM methods.

## B. Justifying grouped sampling and the multichoice method

We illustrated the GSM with an example of biased diffusion. By essentially re-ordering the key steps of the Next Subvolume Method[15], the GSM substantially reduced the sample variance compared with the ungrouped SPM method. This SPM-variant then produced a lower variance than the CPM method. The two-tier simulation procedure of the GSM, meant that, as far as possible, the same ratio of left diffusion to right diffusion events could be maintained in both Systems $X$ and $Y$. By ensuring that the precise location of the diffusing particle is not as important as the direction in which the particle diffuses, a decreased variance was achieved.

The multichoice method is useful for situations where there are substantial qualitative differences in stochastic behaviour in different voxels. With case study II, there system dynamics in front of the wave-front are quite different to the dynamics behind the wave-front. The multichoice method can choose between the SPM and CPM according to the stochastic behaviour of the particular sample path. The multichoice method therefore explicitly accounts for the spatial variation inherent in problems modeled with the RDME by using different coupling methods for the events taking place in different voxels. Equation 12 is written as

$$\boldsymbol{X}(T) = \boldsymbol{X}(0) + \sum_{j=1}^{J} \sum_{\ell=0}^{\infty} \mathcal{P}_{j\ell} \left( \int_{s_\ell}^{s_{\ell+1}} \eta_j^X(\boldsymbol{X}(t)) \mathrm{d}t \right) \cdot \kappa_j,$$

where $\{s_0, \ldots, \}$ represents a mesh of interest, then the multichoice method can essentially be viewed as a method for dynamically choosing a mesh $\{s_0, \ldots, \}$ for each individual event.

## C. Implementation issues

We have presented our results in terms of the number of simulations required to estimate the required sensitivities, and not the overall simulation time. The rationale for this approach is that the performance of the various simulation methods (including the DM, MNRM, CPM, SPM, etc.) can all be implemented with varying degrees of efficiency, and it is not our intention to make such comparisons in this work. All the simulations in this work were produced with `C++` code, according to the `C++11` standard and our results are hardware independent. The SPM is, in the view of the authors, slightly easier to implement than the CPM. Sample code will be available at `http://people.maths.ox.ac.uk/lesterc`.

## D. Outlook

The SPM and CPM can both be implemented to accurately estimate parametric sensitivities of spatially-extended stochastic models. The grouped sampling and multichoice extensions explicitly consider the characteristic dynamics of a spatially-extended network, thereby offering increased efficiency and flexibility. A number of case studies have been considered. Future work should concentrate on rigorously evaluating which parameter sensitivity estimation method should be preferred.